\documentstyle[twocolumn,aps,prb]{revtex}

\newcommand{\half}{\frac{1}{2}}
\begin{document}
\draft
\title{Variational derivation of density functional theory}
\author{R. K. Nesbet}
\address{
IBM Almaden Research Center,
650 Harry Road,
San Jose, CA 95120-6099, USA}
\date{\today}
\maketitle
\begin{center}For {\em Phys.Rev.B} \end{center}
\begin{abstract}
It is shown here that Kohn-Sham equations cannot be derived from
Hohenberg-Kohn theory without an additional postulate.  Assuming that 
a functional derivative with respect to total electron density exists
leads in general to a theory inconsistent with the exclusion principle.
A mathematically and physically correct variational theory of the 
Kohn-Sham model can be developed using functional derivatives with 
respect to orbital densities.  These partial (G\^ateaux) derivatives 
can be constructed explicitly from general N-electron theory and are 
defined throughout the orbital Hilbert space.  This theory is
consistent with the local density approximation (LDA), but does not in 
general imply multiplicative local exchange-correlation potentials.  
Progress beyond the LDA in condensed-matter physics requires development
of methodology for nonlocal exchange and correlation potentials.
\end{abstract}
\section{Introduction}
Hohenberg and Kohn\cite{HAK64} proved, for nondegenerate 
states, that the ground-state total electronic density function
determines the external potential function acting on an interacting
N-electron system.  This is easily extended to a spin-dependent
density, so spin indices and summations are assumed here, but omitted
from the notation.  Implementation of the implied density-functional 
theory (DFT) followed only
after introduction of the orbital model of Kohn and Sham\cite{KAS65},
who applied the Hohenberg-Kohn theory of the universal ground-state
functional $F_s=E-V$ to the kinetic energy functional $T_s$, to 
which $F_s$ reduces for noninteracting electrons.  For noninteracting
electrons, the density function $\rho({\bf r})$ is a sum of orbital
densities $\rho_i=\phi^*_i\phi_i$, one for each occupied orbital 
function of the Kohn-Sham model state.  DFT is developed in many
review articles and monographs, exemplified by \cite{PAY89,DAG90}.
\par It is widely assumed that Hohenberg-Kohn theory implies the
existence of an exact theory that can be represented by the Kohn-Sham
model, using only multiplicative local potential functions.  It is
shown here that this assumption is based on a misconception of the
implications of Hohenberg-Kohn theory.  If the accepted interpretation
were correct, an exact Thomas-Fermi theory (TFT) would exist, in 
conflict with Kohn-Sham DFT and with the exclusion principle for 
electrons.  The assumed existence of exact local exchange-correlation
potentials leads to several problems or paradoxes, resolved by the
present analysis.  Recent quantitative tests\cite{CAN01} confirm a  
longstanding result of optimized effective potential (OEP) calculations,
that the best possible local exchange potential does not reproduce
unrestricted Hartree-Fock (UHF) ground-state energies and 
densities\cite{ALT78}.  The exchange-only limit of DFT linear response
theory\cite{PGG96} is inconsistent with the time-dependent Hartree-Fock
theory of Dirac\cite{DIR30}, due to failure of locality\cite{NES99}.  
\par The present analysis has important practical implications for 
computational methodology relevant to the electronic structure of large 
molecules and solids.  In a formally exact theory, deduced directly 
from N-electron theory, nonlocal exchange-correlation potentials can be 
constructed that are free of electronic self-interaction and not 
inherently restricted to weak electronic correlation or to short-range 
interactions\cite{NES01a}.  Progress beyond the local density
approximation requires the timely development of computational 
methods appropriate to nonlocal potentials\cite{NES03a}.

\section{DFT practice versus DFT concepts }
\par Practical applications of DFT use the Kohn-Sham Ansatz,
$\rho=\sum_i n_i\rho_i$, as a postulate for the density.  Occupation
numbers $n_i$ are unity for occupied orbitals.  Kinetic energy is 
defined (in the nonrelativistic limit) by the Schr\"odinger functional
of occupied orbital functions, $T=\sum_i n_i(i|{\hat t}|i)$, where
${\hat t}=-\half\nabla^2$ in Hartree units.
Given an external potential function $v({\bf r})$, the potential
energy $V=\sum_i n_i(i|v|i)$ is an orbital functional, as is the Hartree
energy $E_h$.  By the Kohn-Sham Ansatz, any assumed 
$E_{xc}[\rho]$ is a functional of the occupied orbital functions. 
Given $F=T+E_h+E_{xc}$ as an orbital functional,
derivation of the orbital Euler-Lagrange (OEL) equations 
\begin{eqnarray}\label{KSeqs}
{\cal F}\phi_i=\{\epsilon_i-v({\bf r})\}\phi_i
\end{eqnarray}
is elementary, using Lagrange multipliers $\epsilon_i$ for independent
orbital normalization $(i|i)=1, i\leq N$.  Linear operator ${\cal F}$ 
is defined by the orbital functional derivative
\begin{eqnarray}
{\cal F}\phi_i=\frac{\delta F}{n_i\delta\phi^*_i}.
\end{eqnarray}
$F$ is assumed to be defined such that ${\cal F}$ is hermitian.
For noninteracting electrons, and also in the local-density 
approximation (LDA), when $E_{xc}$ is the integral of a local function
of $\rho$, the OEL equations (\ref{KSeqs}) reduce explicitly to 
Kohn-Sham equations\cite{KAS65} with local potentials.
\par Eqs.(\ref{KSeqs}) imply equivalent equations\cite{NES02} 
\begin{eqnarray}\label{KSden}
\frac{\phi^*_i{\cal F}\phi_i}{\phi^*_i\phi_i}=
v_{Fi}({\bf r})=\epsilon_i-v({\bf r}), 
\end{eqnarray}
to be solved for the orbital density functions $\rho_i$.  The Lagrange 
multipliers $\epsilon_i$ are to be determined so that independent
normalization conditions $(i|i)=\int\rho_i d^3{\bf r}=1$ are satisfied.
These modified Thomas-Fermi equations are operationally equivalent
to the Kohn-Sham or OEL equations, consistent with the exclusion 
principle.
\par This practical derivation makes no use of Hohenberg-Kohn theory
except to justify particular approximations to $E_{xc}$.  The density 
function is constructed after the fact from solutions of the variational
equations.  To define the model state, some rule $\Psi\to\Phi$ must
be postulated to supplement the Kohn-Sham Ansatz for the density.
Examples are $\rho_{\Phi}=\rho_{\Psi}$ (Kohn-Sham),
$\Psi=\Phi$ (unrestricted Hartree-Fock), and $(\delta\Phi|\Psi)=0$,
the Brueckner-Brenig condition that selects the model state of greatest
weight in $\Psi$\cite{BAW56,BRE57}.  A derivation starting from 
Hohenberg-Kohn theory is more problematic, since a formal theory of
allowable densities is required\cite{LIE83}.  In the orbital theory
described above, this is avoided by recognizing that
densities freely constructed from orbital functions in the
usual Hilbert space suffice for a valid N-electron theory.  

\section{Incompleteness of Hohenberg-Kohn theory}
\par The Hohenberg-Kohn ground-state functional $F_s$ is defined only 
for normalized total densities, $\int\rho=N$.  Standard variational 
theory constrains normalization using the modified Lagrange 
functional $F_s+V-\mu[\int\rho-N]$, which is to made stationary by
varying $\rho$ with unconstrained normalization\cite{NES03}.  $\mu$ is 
adjusted to satisfy the normalization constraint.  This requires the 
definition of $F_s$ to be extended into all infinitesimal neighborhoods
of normalized densities consistent with the orbital Hilbert space.
Without such an extended definition, Hohenberg-Kohn theory is 
incomplete: it cannot determine the variational Euler-Lagrange equations
for $\rho$. 
\par This point is evident in the rigorous derivation of Englisch and
Englisch\cite{EAE84}.  In their Eq.(4.1) they propose to complete their
derivation by inserting a single constant, which would play the role
of the usual chemical potential $\mu$ if the implied Euler-Lagrange
equations were valid.  However, comparison with Eqs.(\ref{KSden}),
implied by physically correct Kohn-Sham equations,  indicates that a 
single constant does not suffice to enforce the exclusion
principle\cite{NES98}.  Independent normalization of each orbital
density is required.  The derivation in\cite{EAE84}, restricted to 
normalized total densities, cannot distinguish between one or several
constants, all of which drop out of the variational integrals.  As
shown here, this is a crucial limitation of Hohenberg-Kohn theory.
\par If the functional differential 
\begin{eqnarray}
\delta F_s=\int d^3{\bf r}v_F({\bf r})\delta\rho({\bf r}),
\end{eqnarray}
is uniquely defined for unrestricted variations, 
it determines a total (Fr\'echet\cite{BAB92}) density functional
derivative
\begin{eqnarray}
\frac{\delta F_s}{\delta\rho({\bf r})}=v_F({\bf r}).
\end{eqnarray}
This is stated as the unique definition of a density functional
derivative in standard DFT literature\cite{PAY89,DAG90}.
If such a Fr\'echet derivative exists, varying the Lagrange
functional using standard variational theory\cite{NES03}
implies the Euler-Lagrange equation
\begin{eqnarray}\label{TFeq}
v_F({\bf r})=\mu-v({\bf r}).
\end{eqnarray}
This is a Thomas-Fermi equation, to be solved by varying a trial density
for specified $\mu$, then adjusting $\mu$ to satisfy the
normalization condition.  It is inconsistent with Eqs.(\ref{KSden})
unless all $\epsilon_i$ are equal\cite{NES98,NES02a}.
\par Because Thomas-Fermi theory (TFT) enforces only total density
normalization, it can determine only a single Lagrange multiplier $\mu$.
This does not exclude solutions that violate the exclusion principle.
The example of the lowest $1s2s\:^3S$ state of
an atom containing two noninteracting electrons is discussed below. 
In this example, two independent orbital energies $\epsilon_{1s}$ and
$\epsilon_{2s}$ are required, not just the single parameter $\mu$. 
The conclusion is that the required Fr\'echet density functional
derivative cannot exist in a physically correct theory for any 
compact system with more than one electron of each spin.

\section{The exclusion principle in DFT}
In any independent-electron orbital model, the exclusion principle
requires independent normalization of each orbital density.  This is 
explicit in any physically correct theory of noninteracting electrons,
implicit in the Kohn-Sham equations, and is a standard aspect of
Hartree-Fock equations.  If the Kohn-Sham density decomposition
is added as a postulate to the analysis of Englisch and Englisch,
the constraint of independent orbital normalization requires 
independent variation of the orbital partial densities.  Such density  
variations, driven by free (unnormalized) variations of occupied
orbital functions, determine Eqs.(\ref{KSden}) as Euler-Lagrange
equations for the orbital densities\cite{NES02}.  These equations are 
expressed in terms of partial (G\^ateaux\cite{BAB92}) functional
derivatives of the universal functional F=E-V.  Independent 
normalization of each orbital density introduces independent Lagrange
multipliers, which are the one-electron energies in the corresponding 
orbital Eqs.(\ref{KSeqs}).  
\par Eqs.(\ref{KSden}) and (\ref{TFeq}) can be reconciled only if all
orbital energy eigenvalues are equal\cite{NES98,NES02a}.
This is seen here to be a direct consequence of the different 
normalization constraints.  That independent normalization of the
orbital densities is physically correct, specifically enforcing the
exclusion principle in Hartree-Fock theory and Kohn-Sham DFT, is obvious
in the example of an atom with two noninteracting electrons of parallel
spin.  The ground state is $1s2s\;^3S$.  The density constraints are:
\begin{eqnarray*}
DFT: \int\rho_{1s}=1,\;\int\rho_{2s}&=&1\\
TFT: \int(\rho_{1s}+\rho_{2s})&=&2.
\end{eqnarray*}
This TFT constraint does not exclude the nonphysical solution
\begin{eqnarray*}
TFT: \int\rho_{1s}=2,\;\int\rho_{2s}&=&0,
\end{eqnarray*}
which violates the exclusion principle.

\section{G\^ateaux functional derivatives}
Euler-Lagrange equations are implied for interacting and noninteracting 
electrons, respectively, only if the definitions of 
$F_s$ and $T_s$ can be extended to unnormalized densities generated by
infinitesimal variations unconstrained in the orbital Hilbert space. 
Given $F_s$ as a functional of total density, it is also an orbital 
functional as a result of the Kohn-Sham Ansatz.  
The orbital functional derivative in the OEL Eqs.(\ref{KSeqs}) exists if
\begin{eqnarray}
\delta F_s &=&\sum_in_i\int d^3{\bf r} 
\{\delta\phi^*_i({\bf r}){\cal F}\phi_i({\bf r})+cc\}
\end{eqnarray} 
defines a unique functional of the occupied orbital functions.
For variations about a stationary state, Eqs.(\ref{KSeqs}) imply that
\begin{eqnarray} 
\delta F_s &=&\sum_in_i\int d^3{\bf r}
\{\epsilon_i-v({\bf r})\}\delta\rho_i({\bf r}),
\end{eqnarray}
since $\delta\rho_i=\{\delta\phi^*_i\phi_i+cc\}$.
This proves, by construction, the existence of a G\^ateaux functional
derivative whose value for such variations is 
\begin{eqnarray}
\frac{\delta F_s}{n_i\delta\rho_i}=\epsilon_i-v({\bf r}),
\end{eqnarray}
for $i\leq N$.  A total
(Fr\'echet) derivative is defined only if these partial functional 
derivatives are all equal, which requires all $\epsilon_i$ to be equal.
\par G\^ateaux functional derivatives exist in general
throughout the orbital Hilbert space. They can be constructed from any
well-defined density functional, using the Kohn-Sham orbital 
decomposition of the total model density function.  They define local 
potentials indexed by orbital subshells, in general not a global local
potential unless the orbital index drops out.  For example, for
any orbital functional $F[\{\phi_i\}]$,  the  functional differential
\begin{eqnarray}
\delta F&=&\sum_in_i\int d^3{\bf r}\{\delta\phi^*_i{\cal F}\phi_i+cc\}
\nonumber\\
&=&\sum_in_i\int d^3{\bf r}\frac{\phi^*_i{\cal F}\phi_i}
   {\phi^*_i\phi_i}\delta\rho_i,
\end{eqnarray}
is uniquely defined for $i\leq N$ throughout the orbital Hilbert space 
if the orbital functional derivative ${\cal F}\phi_i$ exists.  This
determines the G\^ateaux derivative
\begin{eqnarray}\label{Gdrv}
\frac{\delta F}{n_i\delta\rho_i}=
\frac{\phi^*_i{\cal F}\phi_i}{\phi^*_i\phi_i}=v_{Fi}({\bf r}), 
\end{eqnarray}
valid for arbitrary variations in the orbital Hilbert space.
This logic validates Eqs.(\ref{KSden}) in a density-based theory.
Because this orbital-indexed multiplicative local potential
is singular at any node of $\phi_i$, these G\^ateaux derivatives can
only be equal for all index values if the nodes exactly cancel.
A Fr\'echet derivative can occur only in such a special case, which is
very difficult to construct unless the functional is defined as in the
local density approximation.
\par Hence extending the definition of $F_s[\rho]$ so as to determine
TFT equations conflicts in general with the exclusion principle.
In contrast, an extended functional $F_s[\{\rho_i\}]$ and its
G\^ateaux derivatives exist, implying Eqs.(\ref{KSden}). 
The corresponding OEL equations reduce to Kohn-Sham equations in the
LDA, but locality of kinetic,
exchange, and correlation potentials is not implied for more general 
energy functionals.  The mathematical statement of this situation is 
that the Hohenberg-Kohn argument does not imply the existence
of a Fr\'echet functional derivative, equivalent to a multiplicative 
local potential function, for any system with more than one Fermi-Dirac 
electron of each spin.  This conclusion does not conflict with the 
rigorous theory of Englisch and Englisch\cite{EAE84}.  Their argument, 
restricted to normalized ground-state total densities, cannot 
distinguish between Fr\'echet and G\^ateaux functional derivatives.

\section{Orbital functionals in an exact theory}
Formally exact orbital functionals for kinetic, exchange,
and correlation energy can be deduced from standard N-electron
theory.  They can be incorporated into a formally exact mean-field
theory in which the OEL equations extend Kohn-Sham theory to
nonlocal potentials (orbital-indexed local potentials).  Progress
beyond the local-density approximation requires development of
computational methodology for such nonlocal potentials. 
\par
The variational theory of independent-electron models is most simply
developed as an orbital functional theory (OFT)\cite{NES01a}. For an
N-electron eigenstate such that $(H-E)\Psi=0$ and any rule $\Psi\to\Phi$
that determines a model or reference state $\Phi$, unsymmetric
normalization $(\Phi|\Psi)=(\Phi|\Phi)=1$ implies that 
$E=(\Phi|H|\Psi)=E_0+E_c$,
where $E_0=(\Phi|H|\Phi)$ is an explicit orbital functional, and 
$E_c=(\Phi|H|\Psi-\Phi)$ defines the correlation energy.
Restricting the discussion to nondegenerate states, $\Phi$ is a Slater 
determinant constructed from occupied orbital functions $\{\phi_i\}$, 
with occupation numbers $n_i=1$.  Spin indices and sums are assumed here
but are suppressed in the notation.  For orthonormal orbital functions, 
$E_0=T+U+V$, where 
\begin{eqnarray}\label{TVops}  
T=\sum_in_i(i|{\hat t}|i);\;V=\sum_in_i(i|v|i),
\end{eqnarray}
for ${\hat t}=-\half\nabla^2$.  Two-electron functionals are defined
for $u=1/r_{12}$ by $U=E_h+E_x$, where
\begin{eqnarray}\label{Uops}
E_h= \half\sum_{i,j}n_in_j(ij|u|ij);\;
E_x=-\half\sum_{i,j}n_in_j(ij|u|ji).
\end{eqnarray}
If ${\cal Q}=I-\Phi\Phi^{\dagger}$,
$E_c=(\Phi|H|\Psi-\Phi)=(\Phi|H|{\cal Q}\Psi)$.
This implies an exact but implicit orbital functional\cite{NES01a} 
\begin{eqnarray}\label{Ecexp}  
E_c=-(\Phi|H[{\cal Q}(H-E_0-E_c-i\eta){\cal Q}]^{-1}H|\Phi),
\end{eqnarray}
for $\eta\to0+$.  Because $E_c$ is determined by the antisymmetric
model function $\Phi$, its functional derivatives cannot introduce
spurious electronic self-interaction. 
\par Orbital Euler-Lagrange equations follow immediately from
standard variational theory\cite{NES03}, in terms of the 
orbital functional derivatives
\begin{eqnarray}\label{ofds}
\frac{\delta T}{n_i\delta\phi^*_i}={\hat t}\phi_i;\;
\frac{\delta U}{n_i\delta\phi^*_i}={\hat u}\phi_i;\;
\nonumber\\
\frac{\delta V}{n_i\delta\phi^*_i}=v({\bf r})\phi_i;\;
\frac{\delta E_c}{n_i\delta\phi^*_i}={\hat v}_c\phi_i,
\end{eqnarray}
using ${\hat u}=v_h({\bf r})+{\hat v}_x$, where $v_h$ is the classical
Coulomb potential, and ${\hat v}_x$ is the Fock exchange operator.  
Defining $F=E-V=T+U+E_c$, its orbital functional derivative is 
\begin{eqnarray}\label{fdr}
\frac{\delta F}{n_i\delta\phi^*_i}={\cal F}\phi_i=
 \{{\hat t}+{\hat u}+{\hat v}_c\}\phi_i.
\end{eqnarray}
For independent subshell normalization, Lagrange terms 
$\epsilon_i[(i|i)-1]$ are subtracted from the energy functional.  
The variational condition is 
\begin{eqnarray}
\int d^3{\bf r}\sum_in_i(\delta\phi^*_i
 \{{\cal F}+v-\epsilon_i\}\phi_i+cc)=0,
\end{eqnarray}
for unrestricted orbital variations in the usual Hilbert space. 
This implies the OEL equations
\begin{eqnarray}\label{OEL}
{\cal F}\phi_i=\{\epsilon_i-v\}\phi_i,\;\;i=1,\cdots,N.
\end{eqnarray}
\par This theory can be expressed in terms of orbital densities,
using the modified Lagrange functional
$F+V-\sum_i n_i\epsilon_i[\int\rho_i-1]$.  Using Eq.(\ref{Gdrv}),
the stationary condition is expressed as a relationship between 
local potential functions, 
\begin{eqnarray}
\frac{\phi^*_i{\cal F}\phi_i}{\phi^*_i\phi_i}=v_{Fi}({\bf r})
               =\epsilon_i-v({\bf r}),\;\;i=1,\cdots,N.
\end{eqnarray}
These equations reproduce Eqs.(\ref{KSden}), derived directly from the
OEL equations.  This derivation generalizes Eq.(\ref{TFeq}) of TFT to
incorporate the exclusion principle\cite{NES02}.  Orthogonality 
constraints consistent with independent variation of orbital densities
follow from the operationally equivalent OEL equations.

\end{document}